# Exploring non-residential technology adoption: an empirical analysis of factors associated with the adoption of photovoltaic systems by municipal authorities in Germany

Maren Springsklee [a], Fabian Scheller[b] *

[a] Institute for Infrastructure and Resources Management (IIRM), Faculty of Economics and Management Science, Leipzig University, Grimmaische Str. 12, 04109 Leipzig, Germany

[b] Faculty of Business and Engineering, University of Applied Sciences Würzburg-Schweinfurt, Ignaz-Schön-Straße 11, 97421 Schweinfurt, Germany

*fabian.scheller@fhws.de

## Abstract

This research article explores potential influencing factors of solar photovoltaic (PV) system adoption by municipal authorities in Germany in the year 2019. We derive seven hypothesized relationships from the empirical literature on residential PV adoption, organizational technology adoption, and sustainability policy adoption by local governments, and apply a twofold empirical approach to examine them. First, we explore the associations of a set of explanatory variables on the installed capacity of adopter municipalities (N=223) in an OLS model. Second, we use a logit model to analyze whether the identified relationships are also apparent between adopter and non-adopter municipalities (N=423). Our findings suggest that fiscal capacity (measured by per capita debt and per capita tax revenue) and peer effects (measured by the pre-existing installed capacity) are positively associated with both the installed capacity and adoption. Furthermore, we find that institutional capacity (measured by the presence of a municipal utility) and environmental concern (measured by the share of green party votes) are positively associated with municipal PV adoption. Economic factors (measured by solar irradiation) show a significant positive but small effect in both regression models. No evidence was found to support the influence of political will. Results for the role of municipal characteristics are mixed, although the population size was consistently positively associated with municipal PV adoption and installed capacity. Our results support previous studies on PV system adoption determinants and offer a starting point for additional research on non-residential decision-making and PV adoption.

## Highlights

- German Register data is analyzed for PV adoption predictors of municipal authorities.
- Fiscal capacity and peer effect are correlated with the decision and the installed capacity.
- Presence of green voters and municipal utilities are associated with municipal PV adoption.
- Statistically significant but neglectable association of solar radiation level and adoption decision.
- Political will in terms of municipal climate actions show no significant influence.

# 1. Introduction

## 1.1 Problem statement

Solar power constitutes a critical element for transforming the energy production towards renewables and decreasing carbon emissions. Particularly rooftop PV systems provide benefits regarding multiple challenges of the German energy transition: among others, rooftop PV does not exhibit negative trade-offs in land use [1] – as is the case for e.g., bioenergy [2] –, offers high acceptance rates within civil society [3], and is easily scalable due to its modular structure, making it equally suitable for large- and small-scale applications [4]. An analysis commissioned by the German Federal Environment Agency has shown average greenhouse gas reduction potentials for PV electricity for a system operation in Germany between 35 and 57 g CO2-eq./kWh [5]. The deployment of solar power in Germany has increased greatly from 114 Megawatt peak (MWp) in 2000 to a total of 1.8 million PV systems and an accumulated installed capacity of 49.02 Gigawatt peak (GWp) in 2019 [6]. This makes up around 23% of the maximum potential for rooftop PV in Germany, which is estimated at an installed capacity of 210 GWp [7]. To further accelerate the diffusion of PV, some federal states have decided, in various legislative resolutions, to introduce an obligation to install rooftop solar systems [8]. Furthermore, the coalition agreement between the newly-elected governing parties states that "all suitable roof areas will be used for solar energy in the future" [9]. The implementation of PV systems may well become mandatory in the future for both newly constructed and renovated residential and non-residential buildings. The remaining gap in PV coverage at this point would be existing buildings that are not due to be retrofitted.

To determine whether Germany's PV potential can be achieved in the future even without such mandatory PV instalments, it is necessary to understand which factors have driven or inhibited diffusion and adoption in the past. Besides market scale drivers of diffusion – such as lower module costs [10] and feed-in remuneration schemes [11] – research has identified factors that explain the diffusion and adoption of PV via their influence on decision-making processes. Such studies overwhelmingly focus on residential adopters and suggest that factors influencing adoption decisions can include i) socio-economic factors, such as income or education level [4,12–16], ii) psychological factors, including environmental concern and peer effects [4,13–15], iii) techno-economic factors such as solar irradiation, subsidies and electricity cost [4,12,13,16,17], as well as iv) municipal characteristics such as population density and share of detached houses [4,12,13,16]. While numerous studies, therefore, provide empirical evidence on determinants of residential PV adoption [18], little attention has been paid in academic literature to PV diffusion among non-residential building owners such as enterprises and public sector organizations [19].

Public sector organizations consume significant amounts of energy and are displaying an increasing concern with sustainability and the impacts of their consumption behavior on the environment and society as well as economically [20]. Public authorities can support the structural changes necessary for the energy transition either by acting as a diffusion agent, moderator, and coordinator or by providing public network technologies [21–23]. At the municipal level, authorities can drive the diffusion process of renewable technologies by either setting favorable conditions for private actors or can themselves engage in the adoption process and become a 'prosumer'. While municipalities can influence the diffusion of residential solar PV merely via indirect measures such as information campaigns and financial incentives, they have direct control over PV adoption regarding their own municipal buildings – this process is hereafter termed municipal PV adoption. The implementation of such local climate protection measures can offer multiple benefits for municipal authorities: the potential to save costs, generate new revenue streams and regional added value, gain political recognition and prevent climate damages [24]. A survey by Svara et al. [25] highlights the role of cost savings and additional revenue streams by showing that local U.S. governments most frequently adopted measures yielding a direct financial benefit, such as energy efficiency measures or the adoption of solar PV. Moreover, local governments generally favor the adoption of internal government programs as opposed to community-oriented measures such as recycling programs [26] since they can be implemented with little public involvement and resistance [25]. Furthermore, the development of internal climate action plans requires less fiscal capacity and personnel [27].

Within the German context, addressing the research gap on non-residential and specifically municipal PV adoption appears worthwhile due to the relevance of municipal non-residential buildings. There are around three million non-residential buildings in Germany [28]. While this constitutes around 14% of the total building stock, non-residential buildings account for more than 35% of the overall final energy consumption of the building stock [28]. A large share of the associated potential for energy savings and GHG emission reduction lies at the municipal level, which is responsible for around two-thirds of total energy consumption in the public sector [29]. Municipalities own municipal buildings and infrastructure and are significant energy consumers via their administrative offices, schools, hospitals, and street lighting, among others. Therefore, municipalities form a considerable subset of non-residential building owners. For instance, German municipalities own around 175,000 non-residential buildings, with around 48% of municipal sites serving as educational facilities [28]. Their market share, therefore, offers the potential to create critical mass in favor of renewables [30]. Moreover, municipal PV may also positively influence residential PV adoption as outlined by Scheller et al. [22,23]. Larger PV installations – as they may be found on large municipal administrative buildings – are shown to exert stronger peer effects on residential PV adoption for larger installations [12].

So far, reports addressing municipal PV adoption and more generally municipal climate protection measures in Germany focus on best practices and individual case studies, but provide no comprehensive and quantitative data regarding the adoption of municipal PV [24,31–37]. Research on the role and influencing factors of municipalities as consumers and drivers of renewable technologies thus has the potential to complement research on residential PV adoption and provides a starting point for broader empirical research on non-residential PV adoption.

## 1.2 Research objectives

We address this research gap by conducting a quantitative empirical analysis of PV-uptake by German municipalities in 2019. Our research objective is to identify possible influencing factors of municipal PV uptake, both regarding the adoption decision itself and the size of installations. The following research questions guide the article:

(1) Which factors are associated with a higher installed capacity among municipalities that adopted municipal PV in 2019?
(2) Which factors are associated with a higher probability of a municipality having adopted municipal PV in 2019?

We address these research questions by first identifying relevant influencing factors from related empirical literature and operationalizing them with suitable variables. For this purpose, we created a unique database by combining different secondary data sources containing relevant data at the municipal level. We then employ two distinct regression models to statistically test the effect and significance of the identified and operationalized variables regarding municipal PV adoption. While the first econometric model tests the effects of the explanatory variables on the installed PV capacity of municipal authorities in 2019 with an OLS regression analysis, the second model evaluates the effect of these variables on the adoption itself. For this purpose, we draw a random sample of non-adopter municipalities to test the explanatory variables' effects on the probability of adoption of municipal PV between adopter and non-adopter municipalities in a logit regression model. This methodology serves to evaluate the hypotheses that municipalities with better fiscal and institutional capacity, higher potential economic benefits, as well as stronger environmental concern, political will and peer effects show higher levels of municipal PV adoption and installed capacity.

The article is structured as follows. Section 2 provides the theoretical framework and draws on three literature strands to derive potential influencing factors and develop hypotheses regarding their effect. Section 3 introduces the methodology and empirical models employed. Section 4 then presents and discusses model results. The article concludes with a summary and outlook in Section 5.

# 2. Theoretical framework

## 2.1 Potential drivers and barriers of municipal PV adoption

Studies on the adoption of renewable technologies by municipal authorities – or public organizations in general – are scarce. Therefore, our research builds upon three related strands of literature to explain municipal PV adoption: the diffusion and adoption of PV by private households, technology adoption in organizations, and the diffusion and adoption of sustainability policy innovations across local governments. **Table 1** provides an overview of the factors identified in each body of literature and respective studies attributed to each category. While the following paragraphs explain the literature in more detail, **Figure 1** presents the conceptual model containing the factors and hypotheses derived from literature, their associated context within the Technology-Organization-Environment (TOE) framework [38], and the dependent variables to be explained. TOE theory is used to analyse the adoption of innovations at the organizational level and groups influencing factors according to three different contexts in which an adoption decision is made. The underlying assumption of the stated hypotheses is that the identified factors influence both the adoption of municipal PV and installed capacity. Thus, it is expected that variables associated with a higher installed capacity also increase the probability of adoption per se.

**Table 1.** Overview of relevant strands of literature, identified factors, and references.

| Strand of literature | Identified factors | References |
|---|---|---|
| Residential PV adoption | Socio-economic factors (e.g., income, education, age), economic factors (e.g., feed-in tariff, solar irradiation), household characteristics, built environment characteristics (detached houses), peer effects (active and passive), attitudes and values (e.g., environmental concern, perceived benefits and risks) | [4], [12–17], [39–47] |
| Organizational technology adoption | Perceived benefits, costs, available resources, top management support, organizational culture, environmental concern, competitive pressure, technical expertise | [19], [48–59] |
| Policy innovation diffusion | Fiscal capacity, institutional capacity, political will/local policy priorities, environmental concern, climate risk, emission stress, socio-economic attributes, municipal characteristics | [25], [27], [60–68] |

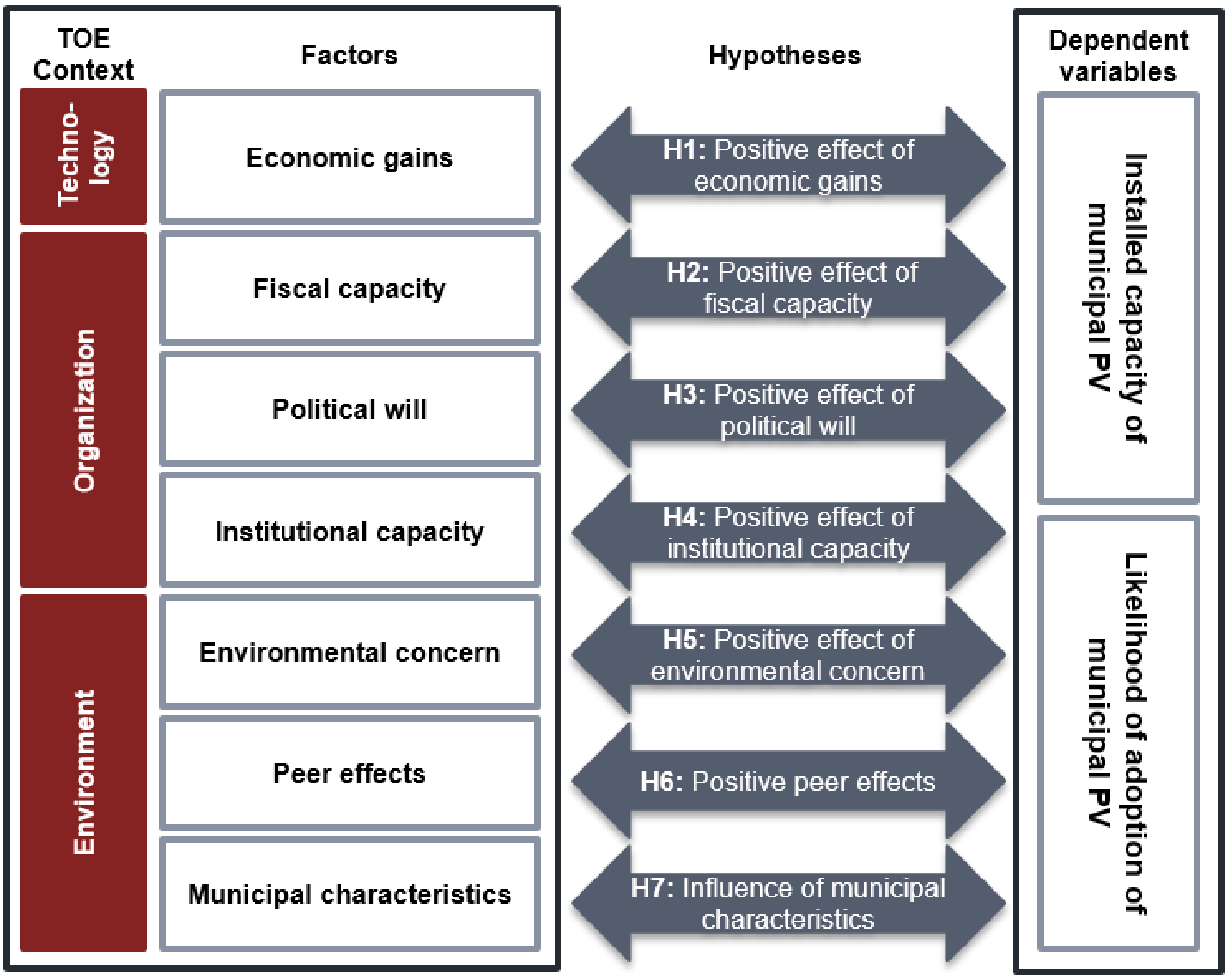

**Figure 1**. Schematic presentation of the conceptual model.

## 2.2 Technology oriented hypotheses

PV promises not only environmental benefits by decreasing emissions of electricity generation but also economic gains to its adopters. These gains can either occur in the form of cost savings via self-consumption – provided the electricity price is higher than the production cost – or remuneration via feed-in tariffs or direct marketing schemes. In their survey of non-residential actors to identify drivers and perceived barriers of PV adoption in Sweden, Reindl and Palm [19] find that economic benefits such as i) cost-saving opportunities, ii) becoming economically sustainable, and iii) protection from electricity price increases, are among the most frequently stated drivers of non-residential PV adoption – along with environmental concerns and technical enablers. Barriers identified in the survey include economic barriers such as taxation or lack of subsidies, regulatory barriers as well as organizational barriers [19]. The amount of solar irradiation provides a well-founded proxy for the profitability of a PV system that captures both the potential benefits of self-consumption and feeding into the grid. For instance. numerous studies find a significant positive effect of solar irradiation on residential PV adoption [12,16,40,46]. Further measures for economic gains include local or state-level financial incentives and subsidies such as feed-in tariffs or tax credits [16,69]. In addition, variables that potentially decrease economic gains (such as PV system costs) as well as variables that increase gains via cost savings (e.g. electricity price from grid, electricity demand) are employed in multiple models explaining residential PV adoption [16,40,46]. Another study uses the net present value (NPV) across five regions to model the relationship between economic benefits and residential PV adoption and finds a positive relation [17]. Overall, the literature provides solid empirical evidence on the positive effect of potential and actual economic gains on PV adoption that is expected to hold true for both residential adopters as well as municipal authorities. This underlying relationship forms the basis of the first hypothesis to be tested in the empirical models.

***H1:*** *Municipalities with the potential for higher economic gains from PV uptake will have higher levels of PV adoption and installed capacity by municipal authorities.*

## 2.3 Organization oriented hypotheses

The fiscal capacity of a municipality can affect the likelihood of adopting innovative technologies [57] and of implementing environmental programs since local governments with higher financial resources are expected to have more funds available for climate action policies and measures [70]. This view is supported by a meta-analysis on climate change policy adoptions by Yeganeh et al. [60], which estimates an increase in fiscal capacity of 1% to be associated with a 0.042% increase in government engagement. Within the federal fiscal context of German municipalities, it is reasonable to hypothesize fiscal capacity to have a significant impact on the climate protection measures such as municipal PV. Since local climate protection measures in Germany are so-called voluntary municipal tasks [71], a municipalities' ability to invest in solar PV on its own building stock is expected to be directly linked to the funds available for such measures. Gonzáles-Limón et al. [62] use three different indicators to measure municipal fiscal health in their study of municipal tax adoptions to promote solar-thermal systems. Specifically, the include per capita (p.c.) income in the municipality, the value of municipal debt p.c., and each municipality's real estate tax rate are included in the analysis. All three variables show a significant positive effect, which is an unexpected result regarding the effect of municipal debt. Although PV promises cost reductions and energy savings in the longer term, upfront investment costs are necessary that vary depending on the PV system size. Therefore, fiscal capacity is expected to influence the installed capacity and adoption of municipal PV positively.

***H2:*** *Municipalities with higher fiscal capacity have higher levels of PV adoption and installed capacity by municipal authorities.*

A further influencing factor may be related to the level of political will authorities display regarding the implementation of climate protection measures. Stronger political will may be linked to higher "local capabilities for information access, technical support and inter-organizational co-ordination" [64]. In a study on sustainability measures in U.S. cities, only 13% of local governments had installed solar panels on government facilities – this activity was pursued only by the most motivated cities in the survey [25]. Tang et al. [66] measure political will as the level of environmental priorities, political commitment, interagency leadership and external campaigns, all of which are self-reported by local planning directors. Their findings suggest that political commitment is highly significant in explaining local planning directors' climate change mitigation actions, including policies addressing renewable and solar energy. Moreover, multiple studies report positive effects of a municipal membership in an intergovernmental climate action network – such as the International Council for Local Environmental Initiatives (ICLEI) or Covenant of Mayors (CoM) – on the implementation of local climate protection measures [62,63]. However, this effect may not hold for residential PV adoption. Kwan [16] tests the hypothesis that membership in the ICLEI positively influences residential adoption due to increased PV-friendly local policies, but finds no significant effect. These findings regarding municipal climate policies indicate that the level of political will in implementing climate protection measures may affect the adoption and installation size of municipal PV.

***H3:*** *Municipalities displaying stronger political will regarding climate protection measures have higher levels of PV adoption and installed capacity by municipal authorities.*

Municipal PV adoption may also be influenced by a municipality's level of institutional capacity, since installing PV systems requires planning. For instance, personnel resources are shown to have a significant positive effect on the implementation of local climate change mitigation and adaptation measures [66]. A further study uses the per capita number of planning professionals as a proxy for a city's technical capacity to implement climate policies and actions and finds a significant positive effect on the adoption of the US Conference of Mayors Climate Protection agreement [70]. The ability of a municipality to implement climate protection measures may also be positively influenced by the presence of a municipal utility [26], since utilities can provide the necessary technical expertise and implementation capacities. This relation is shown to hold for residential PV adoption [43] as well as municipal adoption of climate mitigation policies [72]. Municipalities without municipal utilities may need external technical support in the decision-making process and installation of municipal PV. This is likely to be associated with higher transaction costs, which can include both costs for external technical expertise and a more time-consuming process [73]. This relationship is generally supported by Yeganeh et al. [60], who identify the existence of a sustainability office as well as municipal utility ownership as a measure of government structure. Their results show utility ownership to have a slightly positive average elasticity of around 0.07% regarding the adoption of climate protection measures [60]. Moreover, the presence of a sustainability office displayed an even larger positive effect with an average elasticity of around 0.68%. Based on these results, a positive effect of higher institutional capacity on municipal PV adoption is expected.

***H4:*** *Municipalities with higher institutional capacity have higher levels of PV adoption and installed capacity by municipal authorities.*

Environmental concern is often included as a factor in models explaining residential PV adoption. For instance, Alipour et al. [40] find a significant positive effect of environmental concern on residential PV adoption In their quantitative systematic literature review. The level of environmental concern displayed by citizens may also affect municipal decision making. As Svara et al. [25] argue, policy adoptions are a response to the demands of residents, particularly if they entail high upfront investment costs from new program implementation – which is often the case with sustainability programs. Anderson et al. [61] find positive effects of public opinion shifts towards prioritizing environmental issues on renewable energy policies by European governments. Numerous empirical studies use the share of green party votes as a proxy for the preference for environmental protection – some find evidence of a significant positive effect on climate action and policies [70], while others find no significant effect on residential PV adoption [74]. Increased environmental concern expressed via higher shares of green votes may positively influence local governments in taking climate protection actions and implementing renewable energy policies via the pressure public opinion can exert on governmental decision-making.

***H5:*** *Municipalities in which citizens display greater environmental concern have higher levels of PV adoption and installed capacity by municipal authorities.*

## 2.4 Environment oriented hypotheses

Peer effects constitute a key channel for explaining residential PV diffusion and adoption. They can be grouped into active peer effects operating through word-of-mouth, as well as passive peer effects, which constitute the effect of observing PV systems on different buildings, e.g. around the neighborhood. Research studying the role of such peer effects suggests that active peer effects show a higher effect on adoption decisions [4,45]. Nevertheless, passive peer effects also exert a significant positive effect on adoptions in certain instances and can significantly reduce decision times [75]. In their study on the role of peer effects in residential PV adoption in Sweden, Mundaca and Samahita [13] find that passive peer effects have a stronger influence than active peer effects when the owner of the PV system is someone the adopter knows. Moreover, both Bollinger and Gillingham [15] and Balta-Ozkan et al. [12] find that peer effects are stronger for larger installations. A frequent measure for peer effects is the number of previously existing PV systems within a locality. Numerous empirical results show that the likelihood of adopting PV systems increases with the number of pre-existing installations in spatial proximity due to communication among adopters and potential adopters [4,12,15,45]. The significantly positive peer effects are confirmed across multiple studies within a meta-analysis on determinants of residential PV adoptions [40]. Peer effects measured via the number of pre-existing adoptions are expected to also influence municipal PV adoption positively.

***H6:*** *Municipalities with higher levels of pre-existing adoptions have higher levels of adoption and installed capacity by municipal authorities.*

A range of factors representing municipal characteristics, such as socio-economic factors or the built environment, are included in analyses on climate protection policies and PV adoption as control variables. These factors can include the age and income of inhabitants, population and population density, metropolitan status, and region [25]. Therefore, they capture the environmental context surrounding municipal authorities. The local population is among the most frequently used variables and is shown to have a significant effect on climate action engagement by local governments. For instance, Yeganeh et al. [60] estimate that a 1% increase in population is associated with a 0.623% increase in local government engagement. Population measured as the number of inhabitants of a municipality represents a frequently used control variable in social sciences, as it potentially correlates with socio-economic factors and more generally might "capture factors related to mindset, access to installers, or local peer effects" [4]. Some use population density or share of detached houses as a measure of urbanization [12,76]. While we expect such municipal characteristics to possibly affect municipal PV adoption, no direction of influence is hypothesized due to the lack of a clear transmission channel.

***H7:*** *There is no clear direction of influence of municipal characteristics regarding PV adoption and installed capacity by municipal authorities.*

# 3. Methodology

## 3.1 Research design and estimation methods

We employ two separate regression models to analyze the effect and significance of the identified influencing factors and operationalized variables regarding municipal PV adoption. First, Ordinary Least Squares (OLS) is used to estimate the influence of the factors identified in literature within the group of adopters, i.e., municipalities that installed municipal PV systems in 2019. In a second step, a logistic regression (logit) model is used to estimate the explanatory variables' effect on whether municipalities chose to adopt at all. Here, adopter municipalities are compared to municipalities that did not adopt PV in 2019. This is used to identify whether factors influencing the size of municipal adoption also significantly affect the decision for or against municipal PV. A stratified random sample[1] of non-adopter municipalities is drawn and combined with the sample of adopter municipalities to test the explanatory variables' effects on the probability of adoption of municipal PV.

The factors influencing installed capacity by municipal authorities are analyzed via a cross-sectional multivariate regression model of the general linear form:

$$\boldsymbol{y} = \boldsymbol{X\beta} + \boldsymbol{\varepsilon} \tag{1}$$

where $\boldsymbol{y}$ represents a vector of the dependent variable, $\boldsymbol{X}$ is a matrix of the explanatory variables, $\boldsymbol{\beta}$ a vector of their coefficients, and $\boldsymbol{\varepsilon}$ is a vector of the error terms. The error terms are assumed to be independently identically distributed (i.i.d.) and normally distributed random variables with mean zero and variance $\sigma^2$. The coefficients are estimated using OLS, for which the sum of squared error distances is minimized.[2]

We then use a logistic regression to test the effect of the potential influencing factors on the probability of PV adoption. The probability $\pi_i$ for sample $i$ is calculated in the following manner[3]:

$$\pi_i = P(Y_i = 1) = \frac{e^{\eta_i}}{1 + e^{\eta_i}} \tag{2}$$

[1] To ensure a representative sample of municipalities regarding population sizes, the number of inhabitants serve as strata.
[2] For a general description, see e.g. [77], chapter 3, or [78], chapter 3.
[3] For a general description, see e.g. [77], chapter 17, or [78], chapter 4.

for which

$$\eta_i = \boldsymbol{x_i}\boldsymbol{\beta} \tag{3}$$

where $\eta_i$ describes the linear predictor function, $\boldsymbol{x_i}$ is a vector that includes $k-1$ explanatory variables as well as a constant, and $\boldsymbol{\beta}$ is the corresponding vector of coefficients.

The logit regression equation including the error term $\varepsilon_i$ thus reads:

$$Y_i = P(Y_i = 1) + \varepsilon_i = \pi_i + \varepsilon_i \tag{4}$$

In this model, the dichotomous dependent variable $Y_i$ takes on the value one if a municipality has adopted municipal PV in 2019 (event) and zero for municipalities that did not adopt (non-event). Estimating the coefficients for logistic models is commonly realized with the maximum likelihood estimation (MLE), which produces consistent, efficient, and asymptotically normally distributed estimators as sample sizes increase towards infinity. Generally, MLE works best with balanced samples with an event/non-event distribution close to 50:50 and/or very large samples.

However, using MLE in the case of municipal PV uptake based on a random sample can lead to inefficient results since it can be considered as rare event data. Only 223 of the total 10,799 municipalities in Germany adopted municipal PV in 2019 in our OLS sample, which represents around 2% of all German municipalities. Due to costly and time-consuming data collection, it is unfeasible to draw a random sample that is sufficiently large to adequately capture this distribution. Therefore, drawing a random sample likely results in a highly unbalanced sample with a bias of the MLE towards the non-event [79]. Since most non-events carry little to no additional information of interest and can thus be deemed redundant, a large share of these data points can be removed without a loss of information or consistency in a logit model [80]. This method is known as undersampling: the sample is selected on Y in such a manner that most (or all) event data is included in the sample, while non-event data is undersampled by drawing a smaller random sample [80]. King and Zeng [80] show that in a logit model with undersampling, the maximum likelihood estimator $\hat{\beta}_i$ provides a statistically consistent estimate of $\beta_i$. This means that the coefficient of the logit results can be interpreted without further correction. Only the intercept needs to be corrected if one wants to interpret the probability that an event occurs (see Appendix A.1).

## 3.2 Sample and dependent variable

The dependent variable for this analysis concerns data on PV systems installed by German municipal authorities. Since this information is not readily available, a two-step approach was employed to extract the relevant data. The main database capturing the adoption pattern of PV systems by municipal authorities is the central register for the German power and gas market ('MaStR') introduced by the German Federal Network Agency in 2019 [91]. It is compulsory for all energy market actors[4] to register information regarding their market role as well as the plants they operate. Since the MaStR data alone does not enable an unambiguous assignment and extraction of PV systems installed by municipal authorities, further data is needed for classification. A list of all public sector institutions at the municipal level published by the German federal statistical office [81–83] is compared to the market actor names in the MaStR data to accurately identify municipal authorities that adopted a PV system in 2019. As a result, 232 municipal authorities from 223 different municipalities were identified. The installed capacity of the corresponding 305 municipal PV systems aggregates to 10,212.68 kWp in the year 2019. For these PV systems, the assumption is made that the PV systems are installed on municipal buildings or property.

For the OLS model, the installed capacity of each municipality serves as the dependent variable. Therefore, the OLS sample consists of 223 municipalities. In the logit model, the dependent variable of each of these 223 municipalities is assigned the value one. Additionally, a stratified random sample of 200 municipalities is drawn from all remaining German municipalities, for which the dependent variable takes on the value zero. The random sample is drawn proportionately to the distribution of municipalities regarding their population, as presented in **Table 2**. The overall sample size for the logit regression, therefore, consists of 423 municipalities.

[4] This includes network operators, electricity and gas suppliers, plant operators as well as actors engaging in electricity and gas trading, among others.

**Table 2.** Stratified random sample of non-adopter municipalities.

| Class of inhabitants | Number of municipalities in Germany | Percentage share | Stratified random sample |
|---|---|---|---|
| <2000 | 5666 | 52,5% | 105 |
| 2000-20,000 | 4430 | 41% | 82 |
| 20,000-50,000 | 511 | 4,7% | 9 |
| >50,000 | 192 | 1,8% | 4 |
| Total | 10,799 | 100% | 200 |

## 3.3 Explanatory variables

**Table 3** provides an overview of the explanatory variables and the factors they are attributed to, a description of measurement, predicted relationships as well as data sources. Solar irradiation is used as a measure for economic gains, as this relationship is clearly established in empirical literature and data is available at the municipal level. Municipal tax revenue per capita (p.c.) and municipal debt p.c. are used as a measure of fiscal capacity. Two variables are used as a measure of political will: membership in either the climate action network CoM or ICLEI, as well as whether a municipality has a climate protection strategy in place under the German National Climate Protection Initiative. The variable for climate action membership is included since it is frequently used in empirical literature to operationalize political will. However, Pitt and Bassett [84] suggest that such membership may capture the intention to adopt climate protection measures more than it indicates implementation of actual measures. The variable for climate protection strategy under the National Climate Protection Initiative is a novel measure that is expected to capture political will better for the German context, since applying municipalities are subject to stronger requirements regarding strategy development and receive financial support. Institutional capacity is operationalized by two variables derived from literature: personnel resources and the presence of a municipal utility. The share of green party votes during the most recent national election in Germany in 2017 is used as a measure of environmental concern. Local peer effects are measured as the sum of pre-existing installed capacity adopted within a municipality before 2019. Lastly, two variables are included to operationalize the built environment and socio-economic characteristics of a municipality. House density, measured as the number of residential buildings per square kilometer (km) in a municipality, is used to capture the degree of urbanization, municipal population to capture its size. The natural logarithm was used to transform variables with severely skewed distributions. Per capita measures were used for several variables to ensure comparability between municipalities. All data is collected at the municipal level, with the exception of the variable 'personnel resources'. Since data on public service employees are not available at the municipal level, the share of public sector personnel on the county level divided by the number of inhabitants was used as a proxy.

**Table 3.** Overview of explanatory variables, measures, predicted relationships and data sources.

| Factor | Variable | Variable name | Measures | Predicted relationship | Data sources |
|---|---|---|---|---|---|
| Economic gains | Solar irradiation | *solarirr* | Average annual global solar irradiation in kWh/m² (2005-2016) | + | EC-PVGIS [85] |
| Fiscal capacity | Municipal tax revenue | *lntaxpc* | Natural logarithm of municipal tax revenue per capita in € (2019) | + | GENESIS [86] |
| | Municipal fiscal debt | *lndebtpc* | Natural logarithm of municipal fiscal debt per capita in € (2017) | − | [86] |
| Political will | Climate action network membership | *member* | Membership in Covenant of Mayors (CoM) or ICLEI – Local Governments for Sustainability; 1 in case of membership in either, 0 otherwise | + | CoM [87]/ ICLEI [88] |
| | Climate protection strategy | *strategy* | Integrated climate protection strategy or partial strategy for municipal properties under the national climate protection initiative; 1 if strategy is in place, 0 otherwise | + | National Climate Protection Initiative (NKI) [89] |
| Institutional capacity | Personnel resources | *lnpers* | Natural logarithm of number of public sector employees at the county level/number of inhabitants of municipality (2019) | + | [86] |
| | Municipal utility | *utility* | Presence of municipal utility; 1 if municipal utility exists, 0 otherwise | + | Public sector budgets 2019, Destatis [90] |
| Environmental concern | Green party voting share | *greenvot* | %-share of green party votes during last national election (2017) | + | [86] |
| Peer effects | Pre-existing installations | *lnpeereff* | Natural logarithm of installed capacity (kWp) in municipality up to 2019 (excluding municipal actors) | + | MaStR [91] |
| Municipal characteristics | House density | *lnhousedens* | Natural logarithm of house density (number of residential buildings/municipal area (km²) in 2019) | Control | [86] |
| | Population | *lnpop* | Natural logarithm of population (2019) | Control | [86] |

## 3.4 Model specification and descriptive statistics

The first part of the analysis employs a semi-logarithmic OLS model to investigate the installed capacity of municipal PV across the 223 adopting municipalities. We choose a stepwise approach, testing the model for different specifications to take detect possible multicollinearity between explanatory variables. The full model specification, including all explanatory variables, reads the following:

$$\begin{aligned} lninstcap_i = \beta_0 + \beta_1 solarirr_i + \beta_2 lndebtpc_i + \beta_3 lntaxpc_i + \beta_4 member_i + \beta_5 strategy_i + \beta_6 lnpers_i \\ + \beta_7 utility_i + \beta_8 greenvot_i + \beta_9 lnpeereff_i + \beta_{10} lnhousedens_i + \beta_{11} lnpop_i + u_i \end{aligned} \quad (5)$$

where the dependent variable $lninstcap$ describes the natural logarithm of installed capacity by municipal actors in municipality $i$ in 2019. $\beta_0$ captures the intercept, while $\beta_1$ to $\beta_{11}$ represent the coefficient estimates of the explanatory variables as described in **Table 3**. Lastly, $u_i$ is the error term assumed to be independently and identically distributed with mean zero and constant variance.

The logit model specification includes the same explanatory variables as the OLS model, but the dependent variable $adopt$ is now a binary variable that takes on the value 1 if a municipality adopted municipal PV in 2019 and 0 if it did not. The linear predictor function $\eta_i$ of the logit model is therefore represented by the following term:

$$\begin{aligned} adopt_i = \beta_0 + \beta_1 solarirr_i + \beta_2 lndebtpc_i + \beta_3 lntaxpc_i + \beta_4 member_i + \beta_5 strategy_i + \beta_6 lnpers_i \\ + \beta_7 utility_i + \beta_8 greenvot_i + \beta_9 lnpeereff_i + \beta_{10} lnhousedens_i + \beta_{11} lnpop_i + u_i \end{aligned} \quad (6)$$

Summary statistics are reported in **Table 4** and **Table 5**. Due to a few missing values, the OLS sample reduces to 216 observations. The mean installed capacity of municipal authorities in 2019 was 45.8 kWp, with the smallest PV system installed at a capacity of 2.28 kWp in Bochum and the largest with a capacity of 726 kWp installed in Rott. The mean level of solar irradiation across adopter municipalities is around 1,117 kWh/m²/a, which lies slightly above the overall mean GHI of 1,054 kWh/m²/a for Germany in the period 1981-2010 [92]. The municipalities in the sample have an average level of per capita debt of €1,758 and an average level of per capita tax revenues of €1,482, with both variables showing substantial variation. While 30% of municipalities in the sample have a climate protection strategy in place, only around 3% are members of a sustainability network. Around 40% of municipalities own a municipal utility company. The share of green party votes varies between 2.3% in Trier and 25.5% in Tübingen. The average level of pre-existing installed capacity in the OLS sample is 4,399 kWp, ranging between 24 kWp (min) and 40,479 kWp (max). The correlation matrix is reported in Appendix A.2.

In the logit sample, 52.7% of municipalities had adopted municipal PV, which reflects the undersampling method used. The average solar irradiation of 1,110 kWh/m²/a is marginally lower in the logit sample compared to the OLS sample. Moreover, municipalities in the logit sample indicate both lower levels of per capita debt and per capita tax revenues and slightly lower variation than the adopter sample (OLS). An average of 2.4% of municipalities are members of a sustainability network, while a higher share (21.5%) have a climate protection strategy in place. 27.7% of municipalities in the logit sample own a municipal utility, which is more than 13 percentage points lower than in the OLS sample. The variable capturing personnel resources was excluded in the logit sample due to multicollinearity issues discussed below. The average share of green party votes in the logit sample slightly reduces to 8.2%, compared to 9% in the OLS sample. The average pre-existing installed capacity capturing peer effects is lower in the logit sample and lies at around 3,068 kWp. The OLS sample displays a slightly higher level of house density than the logit sample, with a mean of 98 buildings per km² in the former compared to around 80 buildings per km² in the latter. The minimum and maximum values of the population are equal in both samples, although the average number of inhabitants decreases from around 39,000 in the OLS sample to around 25,000 in the logit sample.

**Table 4.** Descriptive statistics of OLS model variables.

| Variable | Obs | Mean | Std. Dev. | Min | Max |
|---|---|---|---|---|---|
| instcap | 223 | 45.797 | 79.479 | 2.28 | 726 |
| lninstcap | 223 | 3.232 | 1.021 | .824 | 6.588 |
| solarirr | 223 | 1117.417 | 57.299 | 999.29 | 1232.265 |
| debtpc | 216 | 1758.276 | 1605.673 | .37 | 7435.58 |
| lndebtpc | 216 | 6.877 | 1.437 | -.994 | 8.914 |
| taxpc | 220 | 1482.485 | 1748.904 | 578.058 | 25810.842 |
| lntaxpc | 220 | 7.169 | .408 | 6.36 | 10.159 |
| member | 223 | .031 | .175 | 0 | 1 |
| strategy | 223 | .3 | .459 | 0 | 1 |
| personnel | 218 | 1095.246 | 5699.813 | 1.344 | 35745 |
| lnpers | 218 | 5.062 | 1.715 | .295 | 11.216 |
| utility | 223 | .413 | .493 | 0 | 1 |
| greenvot | 220 | .09 | .039 | .023 | .255 |
| peereff | 223 | 4398.899 | 5102.162 | 24 | 40479.302 |
| lnpeereff | 223 | 7.759 | 1.275 | 3.178 | 10.609 |
| housedens | 220 | 98.638 | 89.61 | 6.083 | 461.056 |
| lnhousedens | 220 | 4.173 | .963 | 1.805 | 6.134 |
| pop | 220 | 39429.823 | 125101.71 | 69 | 1484226 |
| lnpop | 220 | 9.202 | 1.575 | 4.234 | 14.21 |

**Table 5.** Descriptive statistics of logit model variables.

| Variable | Obs | Mean | Std. Dev. | Min | Max |
|---|---|---|---|---|---|
| solarpv | 423 | .527 | .5 | 0 | 1 |
| solarirr | 423 | 1110.039 | 57.574 | 986.278 | 1279.51 |
| debtpc | 416 | 1514.807 | 1538.022 | .37 | 7642.21 |
| lndebtpc | 416 | 6.713 | 1.379 | -.994 | 8.941 |
| taxpc | 420 | 1349.261 | 1696.919 | 178.243 | 25810.842 |
| lntaxpc | 420 | 7.047 | .471 | 5.183 | 10.159 |
| member | 423 | .024 | .152 | 0 | 1 |
| strategy | 423 | .215 | .411 | 0 | 1 |
| utility | 423 | .277 | .448 | 0 | 1 |
| greenvot | 420 | .082 | .041 | .009 | .255 |
| peereff | 423 | 3068.447 | 4448.18 | 0 | 40479.302 |
| lnpeereff | 413 | 7.011 | 1.768 | 1.386 | 10.609 |
| housedens | 420 | 79.487 | 81.965 | 6.083 | 461.056 |
| lnhousedens | 420 | 3.92 | .962 | 1.805 | 6.134 |
| pop | 420 | 25037.881 | 96094.453 | 69 | 1484226 |
| lnpop | 420 | 8.496 | 1.675 | 4.234 | 14.21 |

# 4. Results and discussion

## 4.1 OLS regression

Six different model specifications were tested for the OLS model. Regressions were carried out in stages by groups of variables in order to take into account the possible multicollinearity between explanatory variables. The results of the different OLS model specifications are presented in **Table 6**.[5] P-values indicating significance levels are reported at the 1%, 5%, and 10% levels. The F-Test is significant for all model specifications, indicating that the coefficient estimates are jointly significantly distinct from zero (for individual regression tables, see Appendix A.4). This indicates that all specifications offer explanatory power, although to a varying degree.

The variance inflation factor (VIF) was computed after each regression to detect multicollinearity (see Appendix). The highest VIF in specification (2) concerned *lnpop* and was at a still-low level of 3.36. Once *lnpers* was included in specification (3), it increased to 17.27 for *lnpop* and 14.44 for *lnpers*. A VIF above 10 is widely considered an indicator of severe multicollinearity, which is the case here. The construction of the *lnpers* variable explains the heightened VIF. Since data for public sector personnel are not available at the municipal level, the county-level data was used and set in proportion to the share of inhabitants a municipality has within the respective county. The underlying assumption is that a larger share of public sector personnel within a county is allocated to larger municipalities. The measure thus includes the data on municipal population, which explains multicollinearity. Therefore, *lnpers* was excluded in specifications (4) and (5). Specification (6) then tested a model specification that included *lnpers* instead of *lnpop*. This model was estimated to test whether *lnpop* might actually exert its effect via *lnpers*. The intuition behind this is that the positive effect of population on municipal PV installed capacity may be due to the higher amount of personnel resources larger municipalities have available.

The adjusted R-Squared and Bayesian information criterion (BIC) were consulted to assess model fit. Specification (1) displays the lowest BIC of 629.19. However, in combination with the adjusted R-squared, (2) may be an even better fit. It has higher BIC, but its higher adjusted R-squared indicates that adding the variables *member* and *strategy* explains a better portion of the variation. Specifications (3)-(6) show varying levels of explained variation and higher BICs between 642.90 and 647.94, suggesting a poorer model fit compared to (1) and (2). The drop in the adjusted R-squared from (5) to (6) suggests that including *lnpers* in the model explains less of the variation than *lnpop*. Across all specifications, (5) can explain the highest amount variation in the model (Adj. $R^2 = 0.0722$), but also has the highest BIC value, suggesting a poorer model fit than the other specifications.

Solar irradiation shows the expected positive effect and is significant across specifications (1)-(4), but not for (5) and (6). The OLS results thus offer some support of **H1**. However, the effect on the installed capacity is quite small, and therefore, while statistically significant, solar irradiation as a measure of economic gains may not be economically relevant. The first fiscal capacity variable, *lndebtpc*, has a significant negative effect that is consistent over specifications (1)-(5). The results indicate that in 2019, municipalities with higher levels of debt p.c. installed lower capacity of PV than municipalities with lower levels of debt p.c. This finding supports **H2** that fiscal capacity is an influencing factor for explaining the installed capacity of municipal PV adoption. The second variable included to capture the fiscal capacity of a municipality, *lntaxpc,* shows no significance in any of the six specifications and has an unexpected negative sign. The variables used to measure political will (*member*, *strategy*), the institutional capacity variables (*lnpers*, *utility*) show no significance in any of the specifications. Neither does the measure of environmental concern (*greenvot*) or house density (*lnhousedens*). Overall, the OLS results, therefore, do not support **H3-H5**, which hypothesized a positive effect of political will, institutional capacity, and environmental concern on installed capacity of municipal PV. The measure of peer effects, which is the natural logarithm of pre-existing installed capacity, shows a significant and positive effect on municipal installed capacity in 2019. In line with previous findings on the importance of active and passive peer effects as transmission channels, the results, therefore, support **H6**. Lastly, all specifications including *lnpop* showed a positive and significant effect of the variable. This suggests that larger municipalities with more inhabitants tend to install a higher capacity of PV.

[5] The Table reports unstandardized coefficients, which give the percentage change in the explanatory variable associated with a one unit increase (1% increase, for logarithmized variables) of the respective explanatory variable, while holding all others constant.

**Table 6.** Regression analysis results of the different OLS model specifications.

| Variable | (1) | (2) | (3) | (4) | (5) | (6) |
|---|---|---|---|---|---|---|
| **1. Economic gains** | | | | | | |
| solarirr | 0.0027** | 0.0027** | 0.0028** | 0.0025* | 0.0018 | 0.0017 |
| | (0.00) | (0.00) | (0.00) | (0.00) | (0.00) | (0.00) |
| **2. Fiscal capacity** | | | | | | |
| lndebtpc | -0.0997* | -0.1130** | -0.1058* | -0.1101* | -0.1010* | -0.0838 |
| | (0.06) | (0.06) | (0.06) | (0.06) | (0.06) | (0.06) |
| lntaxpc | -0.2289 | -0.2162 | -0.2443 | -0.2234 | -0.2157 | -0.2022 |
| | (0.19) | (0.19) | (0.20) | (0.2) | (0.19) | (0.19) |
| **3. Political will** | | | | | | |
| member | | 0.47133 | 0.4580 | 0.4086 | 0.3989 | 0.4628 |
| | | (0.42) | (0.43) | (0.43) | (0.43) | (0.43) |
| strategy | | 0.2311 | 0.2182 | 0.2203 | 0.1756 | 0.1831 |
| | | (0.17) | (0.17) | (0.17) | (0.17) | (0.17) |
| **4. Institutional capacity** | | | | | | |
| lnpers | | | -0.0468 | | | 0.0750 |
| | | | (0.16) | | | (0.08) |
| utility | | | -0.0334 | -0.0717 | -0.1481 | -0.1114 |
| | | | (0.17) | (0.18) | (0.18) | (0.18) |
| **5. Environmental concern** | | | | | | |
| greenvot | | | | 1.3210 | | 1.5096 |
| | | | | (2.01) | | (2.03) |
| **6. Peer effects** | | | | | | |
| lnpeereff | | | | | 0.1434** | 0.1542** |
| | | | | | (0.06) | (0.07) |
| **7. Municipal characteristics** | | | | | | |
| lnhousedens | -0.0744 | -0.0728 | -0.0562 | -0.0896 | -0.0208 | -0.0404 |
| | (0.11) | (0.11) | (0.12) | (0.12) | (0.12) | (0.12) |
| lnpop | 0.2498*** | 0.20433** | 0.2542** | 0.2192** | 0.1399* | |
| | (0.08) | (0.08) | (0.19) | (0.09) | (0.09) | |
| Constant | 0.5432 | 0.8328 | 0.6159 | 0.9996 | 0.9778 | 1.470 |
| | (1.82) | (1.79) | (2.00) | (1.86) | (1.85) | (1.85) |
| N | 213 | 213 | 211 | 213 | 213 | 211 |
| Adj. R-Sqr | 0.0597 | 0.0635 | 0.0532 | 0.0566 | 0.0722 | 0.0653 |
| BIC | 629.19 | 637.00 | 642.90 | 647.19 | 647.94 | 644.49 |

Note: Standard errors are reported in parentheses, significance levels: *p<0.10, **p<0.05, ***p<0.01

To summarize, solar irradiation, per capita debt, population, and pre-existing installed capacity as a measure of peer effects appear as significant factor of municipal installed capacity in 2019 across most model specifications. Though their significance level changes slightly, their standard errors and magnitude appear robust across specifications. For instance, *lndebtpc* exhibits a coefficient estimate between -0.098 and -0.110 with a robust standard error of 0.06 sd across all models. The OLS results, therefore, suggest that economic factors, fiscal capacity, and peer effects play a role in municipal PV adoption regarding the installed capacity. However, since the OLS model only focuses on adopter municipalities, no conclusion can be drawn whether this also affects municipal PV adoption per se. These effects are instead addressed in the logit model, whose results are presented in the following section.

## 4.2 Logit Regression

Due to multicollinearity issues, *lnpers* is excluded from the logit model. Two model specifications are tested and compared (see **Table 7**, full regression tables are reported in Appendix A.4). Specification (1) includes all variables, while (2) contains all but *lnpeereff*. Results indicate that including *lnpeereff* changes the reported effects of some of the further variables. The likelihood ratio chi-square of the specifications of 140.09 (p-value<0.000) and 131.68 (p-value<0.000), respectively, indicates that the overall model is statistically significant in both cases. Due to the undersampling technique employed, the constant needs to be adjusted by -4.01, which results in a constant of -16.057 in (1) and -17.210 in (2) (see Appendix A.1).

**Table 7.** Regression results of the logit model.

| Variable | (1) | (2) |
|---|---|---|
| **1. Economic gains** | | |
| solarirr | 0.0016 | 0.0040* |
| | (0.00) | (0.00) |
| **2. Fiscal capacity** | | |
| lndebtpc | -0.0918 | -0.1180 |
| | (0.10) | (0.09) |
| lntaxpc | 0.5124* | 0.4808* |
| | (0.31) | (0.29) |
| **3. Political will** | | |
| member | -2.1473** | -2.3526*** |
| | (0.88) | (0.85) |
| strategy | 0.0535 | 0.1941 |
| | (0.33) | (0.32) |
| **4. Institutional capacity** | | |
| utility | 0.2480 | 0.5528* |
| | (0.34) | (0.33) |
| **5. Environmental concern** | | |
| greenvot | 5.6117* | 6.0497* |
| | (3.4) | (3.32) |
| **6. Peer effects** | | |
| lnpeereff | 0.3688*** | |
| | (0.11) | |
| **7. Municipal characteristics** | | |
| lnhousedens | -0.2726 | -0.5209*** |
| | (0.20) | (0.18) |
| lnpop | 0.6270*** | 0.9008** |
| | (0.17) | (0.14) |
| Constant | -12.047*** | -13.2008*** |
| | (3.06) | (2.92) |
| N | 403 | 413 |
| BIC | 483.26 | 500.68 |

Note: Standard errors are reported in parentheses, significance levels: *p<0.10, **p<0.05, ***p<0.01

The logit results suggest that the average level of solar irradiation in a municipality has a significant positive but weak effect on the probability that a municipality is a PV adopter according to the second model specification. This means that higher average solar irradiation and the resulting higher average economic benefit of municipal PV may positively influence its adoption. This result is, however, not robust, since the effect becomes insignificant when *lnpeereff* is included. Therefore, the influence of economic factors hypothesized in **H1** is only somewhat supported by the logit model.

The results of the logit regression analysis do not show a significant effect between the natural logarithm of debt per capita and the probability of municipal PV adoption, although the coefficient has the expected negative sign.

The second fiscal capacity variable, *lntaxpc* is positive and significant at the 10%-level in both model specifications. This indicates that municipalities with higher per capita tax revenues have a higher probability of having adopted municipal PV in 2019. The finding, therefore, supports **H2**.

The unexpected negative significant influence of *member* indicates that membership in a sustainability network decreases the probability of municipal PV adoption in 2019. However, since the overall number of municipalities that are members in either the ICLEI or CoM is quite low (10 municipalities), the variable may not be an adequate measure of political will. The *strategy* variable indicating whether a municipality has a climate protection strategy in place is not significant, but shows the expected positive sign. Overall, no evidence was found in support of **H3**, which hypothesized a positive effect of political will on municipal PV adoption.

Regarding the effect of institutional capacity, the presence of a municipal utility shows a significant positive effect in specification (2), suggesting that the institutional capacity and technical expertise offered by a municipal utility positively influences the likelihood of municipal PV adoption and thus supports **H4**.

The share of green party votes during the national election 2017 shows a significant positive effect on the probability of adoption in 2019. This supports the hypothesis that municipal authorities in municipalities where citizens have a higher level of environmental concern are positively affected in their decision to adopt PV systems (**H5**).

The variable used to operationalize peer effects has a highly significant positive effect on the log-odds of municipal PV adoption in 2019, which supports **H6**. However, when *lnpeereff* is included in the model, the results of *solarirr*, *utility*, and *lnhousedens* become insignificant, suggesting that the results of these variables are not robust. A possible explanation may be that these variables remain quite constant across time and their past values have had an influence on past adoptions by residential actors as well, which is captured by the peer effects variable.

*Lnhousedens* shows a negative and highly significant effect in specification (2). This suggests that municipalities with a higher building density, i.e. urban areas, have a decreased probability of municipal PV adoption. This somewhat stands in contradiction to the significant positive effect of *lnpop* in both logit specifications, as population size may be viewed as another indicator of larger, urban municipalities.

## 4.3 Discussion

The OLS and logit regression analyses provide varying degrees of evidence on all hypotheses but **H3** (political will). Both empirical models reveal that economic gains measured by solar irradiation intensity in a municipality are positively associated with municipal PV adoption and the corresponding installed capacity. However, this effect appears rather small and is not significant across every model specification. The level of debt in a municipality seems to affect the size of the installed PV system by municipal authorities, although no evidence was found regarding whether it affects adoption itself. The upfront investment costs for PV installations are naturally higher, the larger the PV system. Since energy and climate protection-related tasks such as municipal PV are non-compulsory municipal tasks and thus fall within the financial responsibility of a municipality, this effect on installed capacity was to be expected. Although optimally planned and installed systems offer potential for cost reductions in the longer term, these initial investment costs need to be borne. Municipalities with higher fiscal capacity, therefore, appear more likely to have the necessary funds at their disposal, which could translate into higher PV uptake. This is supported by the logit regression, which indicates that municipalities that accrue higher tax revenues p.c. are more likely to have adopted a PV system in 2019. The results are in line with the findings of Yeganeh et al. [60] and Damanpour and Schneider [57], who find a positive effect of fiscal health and municipal wealth, respectively.

Both the OLS and logit model provide no evidence to support the hypothesized effect of political will on municipal PV uptake. This is unexpected, as the role of political will, e.g., measured by membership in a climate action network is frequently stressed in literature [66]. A possible reason for this finding may be that membership in such a network only captures the intention to drive sustainability issues locally, but not the actual implementation of measures [84]. Moreover, the logit model provides weak evidence that institutional capacity measured by the presence of a municipal utility increases the probability of municipal PV adoption. Yet the equivalent could not be confirmed for the OLS model. This suggests that, while institutional capacity may not be associated with higher installed capacity for those municipalities that did adopt, it might affect the decision of whether to adopt. In other words, the presence of a municipal utility may provide the necessary technical expertise for a municipality to install PV in the first place, but once the adoption decision is made, the size of the installed PV system depends on other factors. Additionally, it is shown that environmental concern is positively associated with the probability of municipal PV adoption. This confirms the findings of Wang [70], Alipour et al. [40] and Palm [4] regarding both residential PV adoption and local climate protection policy adoption. Similar to institutional capacity, the differing significance between the OLS and Logit models suggest that environmental concern may affect the initial decision for or against installing municipal PV, but not necessarily the installed capacity itself.

A statistically significant impact of peer effects became apparent during the analysis. Both the OLS and logit results provide evidence of a positive relationship between the pre-existing level of PV in a municipality and the probability of adoption by a municipal authority as well as the installed capacity. This supports the findings of Müller and Rode [76], Balta-Ozkan et al. [12,93], and Dharshing [14], which state that the propensity to install PV is greater, the higher the number of existing systems in local proximity.

The significant negative effect of *lnhousedens* in the second logit specification may capture the effect of detached single-family homes in the following way: municipalities with a higher share of detached houses are likely to have lower levels of house density, as these are more spaced out in comparison to multi-family buildings in urban areas [76]. Since the share of detached houses is shown to increase the adoption of residential PV [12], the same effect might hold for municipal PV adoption. Testing the effect of detached houses on municipal PV adoption would be an interesting extension of the model and offers potential for further investigation.

# 5. Conclusion

We conducted a first cross-sectional exploratory analysis in the field of non-residential PV adoption in Germany by analizing PV uptake of municipal authorities. Due to the novel subject of investigation, our scope was to provide an initial set of quantitative evidence by identifying possible influencing factors as a basis for further study. The empirical analyses confirm that the following factors are positively associated with either municipal PV installed capacity or the likelihood of PV adoption itself: economic factors (measured by solar irradiation); fiscal capacity (measured by debt p.c. and tax revenue p.c.); institutional capacity (measured by presence of a municipal utility); environmental concern (measured by the share of green party votes); and peer effects (measured by the pre-existing installed capacity). No evidence was found to support the influence of political will (measured by climate action network membership and climate protection strategy). Results for the role of municipal characteristics (measured by house density and population) were mixed, although population was consistently shown to be positively correlated with municipal PV adoption and installed capacity. The results indicate that additional measures may be needed to operationalize the identified factors, specifically regarding political will and municipal characteristics of the built environment. Overall, the analysis confirms previous research results regarding non-residential PV adoption to a large extent.

The relationships identified in the OLS and logit models only hold true for the observation year 2019. Further analyses are needed to take various years or even the whole adoption period into account, e.g. by means of a panel analysis. Although this would allow more generalizable results, data availability presents a challenge. Nevertheless, the observed relationships may be a first indication of factors that influence the decision-making of municipal authorities regarding PV uptake. A further limitation of the study results from the fact that some potentially identified and relevant factors of municipal PV adoption could not be included due to data availability. For example, data on municipal energy consumption and the municipal building stock available (and technically feasible) for the installation of PV systems could improve the explanatory power of the developed models. Additionally, a multi-level model may be an appropriate extension to capture the effects of influences of different levels of government. For instance, municipal PV adoption may also be influenced by legislation or support measures, such as subsidies, at the state (Länder) level. Lastly, the measures that were employed in the model may not be the most adequate operationalizations of the described influencing factors. These could therefore be further developed to better reflect the factors they seek to identify. One example could be the peer effects variable, which could be further refined to distinguish between active and passive peer effects, for example, as implemented by Mundaca and Samahita [13]. Further variables that measure economic benefits of PV adoption could be included in a panel model, such as the electricity price, feed-in tariff or PV module costs. As these do not offer enough variation within a cross-sectional analysis, they were not included in the developed empirical models. Nevertheless, the results can further help to understand public organizational decision-making regarding climate protection measures and the adoption of PV by non-residential actors. Understanding what drives voluntary climate action measures such as municipal PV at the local level can potentially support federal and state policymakers in designing policies that can accelerate the rate of PV adoption among non-residential actors, for example by offering an increased compatibility with local incentives and higher cost-effectiveness.

# Acknowledgment

This work was financed by the Saxony State government out of the State budget approved by the Saxony State Parliament within the framework of the project SUSIC no. 100378087.

# Declaration of competing interest

The authors declare that they have no known competing financial interests or personal relationships that could have appeared to influence the work reported in this paper.

# Appendix

## A.1. Undersampling method

In 2019, only 223 of the total 10,799 municipalities in Germany adopted municipal PV, which represents around 2% of all German municipalities. The research subject can accordingly be characterized as rare events data, where the event outcome Y=1 is observed far less frequently than the non-event outcome Y=0. King and Zeng [79] describe this issue in their study examining the probability of a country being in a state of war. According to [79], employing the common procedure of drawing a random sample from the overall population would be problematic, as the probability of drawing an event, i.e. Y=1 is small due to its distribution in the overall population. As a result, the random sample will be unbalanced towards non-events for which Y=0, resulting in a bias of the MLE towards the non-event. Generally, MLE works best with balanced samples with an event/non-event distribution close to 50:50 and/or large samples. King and Zeng [80] argue that in rare events data, the event data is often of interest and thus carries information of higher relevance. They conclude that, since most non-events carry little to no additional information of interest and can thus be deemed redundant, a large share of these data points can be removed without a loss of information or consistency in a logit model. This method is termed undersampling: the sample is selected on Y in such a manner that most (or all) event data is included in the sample, while non-event data is undersampled by drawing a smaller random sample. This way, the sample can be balanced out.

[80] show that in a logit model with undersampling, the maximum likelihood estimator $\hat{\beta}_1$ provides a statistically consistent estimate of $\beta_1$. This means that the coefficient of the logit results can be interpreted without further correction. Only the intercept needs to be corrected if one wants to interpret the probability that an event occurs. Following [80], the prior correction method can be used to correct the estimates for the intercept due to undersampling:

$$\hat{\beta}_0 - ln\left[\left(\frac{1-\tau}{\tau}\right)\left(\frac{\bar{y}}{1-\bar{y}}\right)\right] \qquad \text{Eq. (A.1)}$$

where $\tau$ is the fraction of ones in the population and $\bar{y}$ is the fraction of ones in the sample. Knowledge of the fraction of ones in the overall population is therefore a prerequisite for the prior correction method. In our case, $\tau$ is known and describes the fraction of municipalities that adopted municipal PV in 2019 over all municipalities, i.e. $\frac{223}{10799} = 0{,}02$. $\bar{y}$ is the observed fraction of municipalities that adopted PV in the sample. In our case, an additional random sample of 200 non-adopter municipalities is drawn and added to the sample of 223 adopter municipalities, thus $\bar{y} = \frac{223}{423} = 0{,}53$. To interpret the logit results as probabilities, the intercept would therefore need to be corrected by plugging $\tau$ and $\bar{y}$ into (5), which results in $\hat{\beta}_0 - \ln(55.26) = \hat{\beta}_0 - 4{,}01$. Even without this correction, the coefficient estimates are efficient and can therefore be interpreted in their effective direction, that is, whether they increase or decrease the probability that a municipality has adopted municipal PV in 2019. A negative coefficient then translates into a reduction of the probability and the opposite holds for a positive coefficient.

## A.2. OLS correlation matrix

| Variables | (1) | (2) | (3) | (4) | (5) | (6) | (7) | (8) | (9) | (10) | (11) | (12) |
|---|---|---|---|---|---|---|---|---|---|---|---|---|
| (1) lninstcap | 1 | | | | | | | | | | | |
| (2) solarirr | 0.101 | 1 | | | | | | | | | | |
| (3) lndebtpc | -0.004 | -0.214 | 1 | | | | | | | | | |
| (4) lntaxpc | 0.030 | 0.190 | -0.068 | 1 | | | | | | | | |
| (5) member | 0.135 | -0.116 | 0.214 | 0.062 | 1 | | | | | | | |
| (6) strategy | 0.140 | -0.089 | 0.313 | 0.071 | 0.046 | 1 | | | | | | |
| (7) lnpers | 0.217 | -0.127 | 0.416 | 0.272 | 0.402 | 0.375 | 1 | | | | | |
| (8) utility | 0.081 | -0.172 | 0.415 | 0.081 | 0.217 | 0.252 | 0.532 | 1 | | | | |
| (9) greenvot | 0.118 | 0.222 | 0.103 | 0.169 | 0.222 | 0.150 | 0.252 | 0.276 | 1 | | | |
| (10) lnpeereff | 0.247 | 0.054 | 0.211 | 0.096 | 0.182 | 0.262 | 0.475 | 0.376 | 0.037 | 1 | | |
| (11) lnhousedens | 0.144 | -0.060 | 0.251 | 0.388 | 0.243 | 0.324 | 0.742 | 0.393 | 0.320 | 0.200 | 1 | |
| (12) lnpop | 0.209 | -0.178 | 0.455 | 0.299 | 0.395 | 0.404 | 0.961 | 0.554 | 0.230 | 0.472 | 0.761 | 1 |

## A.3. OLS variance inflation factor

OLS specification 1: Variance inflation factor

| Variable | VIF | 1/VIF |
|---|---|---|
| lnpop | 2.93 | 0.340976 |
| lnhousedens | 2.56 | 0.390872 |
| lndebtpc | 1.35 | 0.738372 |
| lntaxpc | 1.28 | 0.779592 |
| solarirr | 1.12 | 0.894086 |
| Mean VIF | 1.85 | |

OLS specification 2: Variance inflation factor

| Variable | VIF | 1/VIF |
|---|---|---|
| lnpop | 3.36 | 0.297989 |
| lnhousedens | 2.58 | 0.387588 |
| lndebtpc | 1.39 | 0.719381 |
| lntaxpc | 1.28 | 0.778379 |
| strategy | 1.25 | 0.797750 |
| member | 1.22 | 0.817903 |
| solarirr | 1.12 | 0.893288 |
| Mean VIF | 1.74 | |

OLS specification 3: Variance inflation factor

| Variable | VIF | 1/VIF |
|---|---|---|
| lnpop | 17.27 | 0.057907 |
| lnpers | 14.44 | 0.069231 |
| lnhousedens | 2.62 | 0.381859 |
| utility | 1.53 | 0.655461 |
| lndebtpc | 1.46 | 0.683109 |
| lntaxpc | 1.32 | 0.757819 |
| strategy | 1.26 | 0.795727 |
| member | 1.22 | 0.816959 |
| solarirr | 1.15 | 0.868452 |
| Mean VIF | 4.70 | |

OLS specification 4: Variance inflation factor

| Variable | VIF | 1/VIF |
|---|---|---|
| lnpop | 3.78 | 0.264881 |
| lnhousedens | 2.71 | 0.369009 |
| utility | 1.61 | 0.620371 |
| lndebtpc | 1.45 | 0.689796 |
| greenvot | 1.31 | 0.760975 |
| lntaxpc | 1.29 | 0.775340 |
| member | 1.28 | 0.780843 |
| strategy | 1.27 | 0.790270 |
| solarirr | 1.22 | 0.822307 |
| Mean VIF | 1.77 | |

OLS specification 5: Variance inflation factor

| Variable | VIF | 1/VIF |
|---|---|---|
| lnpop | 4.51 | 0.221937 |
| lnhousedens | 2.92 | 0.342334 |
| utility | 1.68 | 0.594865 |
| lnpeereff | 1.54 | 0.647641 |
| lndebtpc | 1.46 | 0.685825 |
| greenvot | 1.34 | 0.748596 |
| lntaxpc | 1.29 | 0.775062 |
| solarirr | 1.29 | 0.776479 |
| strategy | 1.28 | 0.778219 |
| member | 1.28 | 0.780549 |
| Mean VIF | 1.86 | |

OLS specification 6: Variance inflation factor

| Variable | VIF | 1/VIF |
|---|---|---|
| lnpers | 3.76 | 0.265746 |
| lnhousedens | 2.76 | 0.362063 |
| utility | 1.66 | 0.602971 |
| lnpeereff | 1.53 | 0.653212 |
| lndebtpc | 1.42 | 0.705905 |
| greenvot | 1.34 | 0.746360 |
| lntaxpc | 1.28 | 0.783056 |
| strategy | 1.28 | 0.784245 |
| solarirr | 1.27 | 0.788086 |
| member | 1.25 | 0.798771 |
| Mean VIF | 1.75 | |

## A.4. Regression results

OLS specification 1

| Source | SS | df | MS |
|---|---|---|---|
| Model | 18.3493873 | 5 | 3.66987747 |
| Residual | 205.684981 | 207 | .993647254 |
| Total | 224.034369 | 212 | 1.05676589 |

| | | |
|---|---|---|
| Number of obs | = | 213 |
| F(5, 207) | = | 3.69 |
| Prob > F | = | 0.0032 |
| R-squared | = | 0.0819 |
| Adj R-squared | = | 0.0597 |
| Root MSE | = | .99682 |

| lninstcap | Coef. | Std. Err. | t | P>\|t\| | [95% Conf. | Interval] |
|---|---|---|---|---|---|---|
| solarirr | .0026979 | .0012588 | 2.14 | 0.033 | .0002161 | .0051796 |
| lndebtpc | -.0996998 | .0552593 | -1.80 | 0.073 | -.2086429 | .0092433 |
| lntaxpc | -.2289394 | .1944864 | -1.18 | 0.240 | -.6123675 | .1544888 |
| lnhousedens | -.0743658 | .1147074 | -0.65 | 0.518 | -.3005104 | .1517789 |
| lnpop | .2498189 | .0760959 | 3.28 | 0.001 | .0997965 | .3998413 |
| _cons | .5431739 | 1.821341 | 0.30 | 0.766 | -3.047582 | 4.133929 |

OLS specification 2

| Source | SS | df | MS |
|---|---|---|---|
| Model | 21.146173 | 7 | 3.02088186 |
| Residual | 202.888196 | 205 | .989698516 |
| Total | 224.034369 | 212 | 1.05676589 |

| | | |
|---|---|---|
| Number of obs | = | 213 |
| F(7, 205) | = | 3.05 |
| Prob > F | = | 0.0045 |
| R-squared | = | 0.0944 |
| Adj R-squared | = | 0.0635 |
| Root MSE | = | .99484 |

| lninstcap | Coef. | Std. Err. | t | P>\|t\| | [95% Conf. | Interval] |
|---|---|---|---|---|---|---|
| solarirr | .0027321 | .0012569 | 2.17 | 0.031 | .0002541 | .0052101 |
| lndebtpc | -.1130403 | .0558725 | -2.02 | 0.044 | -.2231987 | -.0028818 |
| lntaxpc | -.2161915 | .1942508 | -1.11 | 0.267 | -.5991771 | .1667941 |
| member | .4713306 | .4227735 | 1.11 | 0.266 | -.3622111 | 1.304872 |
| strategy | .231138 | .1650354 | 1.40 | 0.163 | -.0942464 | .5565224 |
| lnhousedens | -.0727644 | .1149633 | -0.63 | 0.527 | -.2994264 | .1538975 |
| lnpop | .2043329 | .0812379 | 2.52 | 0.013 | .044164 | .3645017 |
| _cons | .8328219 | 1.825874 | 0.46 | 0.649 | -2.767078 | 4.432721 |

OLS specification 3

| Source | SS | df | MS |
|---|---|---|---|
| Model | 20.8854579 | 9 | 2.32060644 |
| Residual | 201.800527 | 201 | 1.00398272 |
| Total | 222.685985 | 210 | 1.06040945 |

| | | |
|---|---|---|
| Number of obs | = | 211 |
| F(9, 201) | = | 2.31 |
| Prob > F | = | 0.0171 |
| R-squared | = | 0.0938 |
| Adj R-squared | = | 0.0532 |
| Root MSE | = | 1.002 |

| lninstcap | Coef. | Std. Err. | t | P>\|t\| | [95% Conf. | Interval] |
|---|---|---|---|---|---|---|
| solarirr | .0028183 | .0012843 | 2.19 | 0.029 | .0002859 | .0053507 |
| lndebtpc | -.1058089 | .0587105 | -1.80 | 0.073 | -.2215764 | .0099585 |
| lntaxpc | -.2443104 | .199239 | -1.23 | 0.222 | -.6371771 | .1485563 |
| member | .4580306 | .4261281 | 1.07 | 0.284 | -.3822242 | 1.298286 |
| strategy | .2181518 | .1667886 | 1.31 | 0.192 | -.110728 | .5470316 |
| lnpers | -.0468101 | .1580612 | -0.30 | 0.767 | -.3584809 | .2648606 |
| utility | -.0333629 | .1725266 | -0.19 | 0.847 | -.3735572 | .3068313 |
| lnhousedens | -.056185 | .1172737 | -0.48 | 0.632 | -.2874296 | .1750596 |
| lnpop | .2542225 | .1857561 | 1.37 | 0.173 | -.1120582 | .6205033 |
| _cons | .6159066 | 1.997855 | 0.31 | 0.758 | -3.323537 | 4.55535 |

OLS specification 4

| Source | SS | df | MS |
|---|---|---|---|
| Model | 21.6489262 | 9 | 2.40543625 |
| Residual | 202.385443 | 203 | .996972624 |
| Total | 224.034369 | 212 | 1.05676589 |

| | | |
|---|---|---|
| Number of obs | = | 213 |
| F(9, 203) | = | 2.41 |
| Prob > F | = | 0.0127 |
| R-squared | = | 0.0966 |
| Adj R-squared | = | 0.0566 |
| Root MSE | = | .99849 |

| lninstcap | Coef. | Std. Err. | t | P>\|t\| | [95% Conf. | Interval] |
|---|---|---|---|---|---|---|
| solarirr | .0024763 | .0013148 | 1.88 | 0.061 | -.0001161 | .0050687 |
| lndebtpc | -.1101031 | .0572674 | -1.92 | 0.056 | -.2230184 | .0028122 |
| lntaxpc | -.2234403 | .1953451 | -1.14 | 0.254 | -.6086059 | .1617252 |
| member | .4086084 | .434277 | 0.94 | 0.348 | -.4476638 | 1.264881 |
| strategy | .2202687 | .1664229 | 1.32 | 0.187 | -.1078705 | .5484079 |
| utility | -.0717142 | .1758457 | -0.41 | 0.684 | -.4184325 | .2750042 |
| greenvot | 1.320952 | 2.00933 | 0.66 | 0.512 | -2.640883 | 5.282786 |
| lnhousedens | -.0896217 | .1182541 | -0.76 | 0.449 | -.3227856 | .1435421 |
| lnpop | .2192159 | .0864814 | 2.53 | 0.012 | .0486988 | .389733 |
| _cons | .9995998 | 1.861019 | 0.54 | 0.592 | -2.669806 | 4.669006 |

OLS specification 5

| Source | SS | df | MS | | | |
|---|---|---|---|---|---|---|
| Model | 25.9876013 | 10 | 2.59876013 | Number of obs | = | 213 |
| Residual | 198.046768 | 202 | .980429542 | F(10, 202) | = | 2.65 |
| | | | | Prob > F | = | 0.0047 |
| Total | 224.034369 | 212 | 1.05676589 | R-squared | = | 0.1160 |
| | | | | Adj R-squared | = | 0.0722 |
| | | | | Root MSE | = | .99017 |

| lninstcap | Coef. | Std. Err. | t | P>\|t\| | [95% Conf. | Interval] |
|---|---|---|---|---|---|---|
| solarirr | .0018099 | .0013418 | 1.35 | 0.179 | -.0008357 | .0044556 |
| lndebtpc | -.1010127 | .0569545 | -1.77 | 0.078 | -.2133143 | .0112889 |
| lntaxpc | -.215718 | .1937524 | -1.11 | 0.267 | -.5977545 | .1663185 |
| member | .3910124 | .4307402 | 0.91 | 0.365 | -.4583113 | 1.240336 |
| strategy | .1770665 | .1663093 | 1.06 | 0.288 | -.1508583 | .5049914 |
| utility | -.1476731 | .1780799 | -0.83 | 0.408 | -.498807 | .2034608 |
| greenvot | 1.859979 | 2.008998 | 0.93 | 0.356 | -2.101317 | 5.821275 |
| lnpeereff | .1403233 | .0667051 | 2.10 | 0.037 | .0087956 | .271851 |
| lnhousedens | -.02076 | .121752 | -0.17 | 0.865 | -.2608278 | .2193079 |
| lnpop | .1398562 | .0936916 | 1.49 | 0.137 | -.0448828 | .3245951 |
| _cons | .9778498 | 1.845543 | 0.53 | 0.597 | -2.66115 | 4.61685 |

OLS specification 6

| Source | SS | df | MS | | | |
|---|---|---|---|---|---|---|
| Model | 24.4569353 | 10 | 2.44569353 | Number of obs | = | 211 |
| Residual | 198.22905 | 200 | .991145248 | F(10, 200) | = | 2.47 |
| | | | | Prob > F | = | 0.0084 |
| Total | 222.685985 | 210 | 1.06040945 | R-squared | = | 0.1098 |
| | | | | Adj R-squared | = | 0.0653 |
| | | | | Root MSE | = | .99556 |

| lninstcap | Coef. | Std. Err. | t | P>\|t\| | [95% Conf. | Interval] |
|---|---|---|---|---|---|---|
| solarirr | .0016811 | .0013395 | 1.25 | 0.211 | -.0009603 | .0043225 |
| lndebtpc | -.0838283 | .0573843 | -1.46 | 0.146 | -.1969843 | .0293276 |
| lntaxpc | -.2021921 | .194745 | -1.04 | 0.300 | -.586209 | .1818248 |
| member | .4628358 | .428188 | 1.08 | 0.281 | -.3815066 | 1.307178 |
| strategy | .1831713 | .1669275 | 1.10 | 0.274 | -.1459924 | .5123351 |
| lnpers | .0750423 | .080158 | 0.94 | 0.350 | -.083021 | .2331057 |
| utility | -.1114184 | .1787256 | -0.62 | 0.534 | -.4638467 | .2410099 |
| greenvot | 1.509636 | 2.028167 | 0.74 | 0.458 | -2.489698 | 5.50897 |
| lnpeereff | .1541683 | .0668035 | 2.31 | 0.022 | .0224388 | .2858978 |
| lnhousedens | .0403768 | .1196646 | 0.34 | 0.736 | -.1955894 | .2763431 |
| _cons | 1.470393 | 1.851062 | 0.79 | 0.428 | -2.179709 | 5.120495 |

Logit specification 1 (with *lnpeereff*)

| Logistic regression | | Number of obs | = | 403 |
|---|---|---|---|---|
| | | LR chi2(10) | = | 140.09 |
| | | Prob > chi2 | = | 0.0000 |
| Log likelihood = | -208.6342 | Pseudo R2 | = | 0.2514 |

| solarpv | Coef. | Std. Err. | z | P>\|z\| | [95% Conf. | Interval] |
|---|---|---|---|---|---|---|
| solarirr | .0016226 | .0022447 | 0.72 | 0.470 | -.0027769 | .0060221 |
| lndebtpc | -.0918016 | .0998866 | -0.92 | 0.358 | -.2875756 | .1039725 |
| lntaxpc | .5123633 | .3071281 | 1.67 | 0.095 | -.0895967 | 1.114323 |
| member | -2.147372 | .8779456 | -2.45 | 0.014 | -3.868114 | -.4266307 |
| strategy | .0534878 | .3345871 | 0.16 | 0.873 | -.6022909 | .7092665 |
| utility | .2480741 | .3414382 | 0.73 | 0.467 | -.4211326 | .9172808 |
| greenvot | 5.6117 | 3.398214 | 1.65 | 0.099 | -1.048678 | 12.27208 |
| lnpeer | .3687995 | .108043 | 3.41 | 0.001 | .1570391 | .5805598 |
| lnhouse | -.2725977 | .2027278 | -1.34 | 0.179 | -.6699368 | .1247414 |
| lnpop | .6270064 | .1668181 | 3.76 | 0.000 | .3000489 | .9539638 |
| _cons | -12.04772 | 3.060485 | -3.94 | 0.000 | -18.04616 | -6.049278 |

Logit specification 2 (without *lnpeereff*)

| Logistic regression | | Number of obs | = | 413 |
|---|---|---|---|---|
| | | LR chi2(9) | = | 131.68 |
| | | Prob > chi2 | = | 0.0000 |
| Log likelihood = | -220.22353 | Pseudo R2 | = | 0.2302 |

| solarpv | Coef. | Std. Err. | z | P>\|z\| | [95% Conf. | Interval] |
|---|---|---|---|---|---|---|
| solarirr | .0039849 | .0021047 | 1.89 | 0.058 | -.0001402 | .00811 |
| lndebtpc | -.1179715 | .0939846 | -1.26 | 0.209 | -.3021779 | .0662349 |
| lntaxpc | .4807738 | .288432 | 1.67 | 0.096 | -.0845425 | 1.04609 |
| member | -2.352648 | .8524359 | -2.76 | 0.006 | -4.023391 | -.681904 |
| strategy | .1941192 | .3239906 | 0.60 | 0.549 | -.4408907 | .829129 |
| utility | .5528247 | .3313039 | 1.67 | 0.095 | -.096519 | 1.202168 |
| greenvot | 6.049655 | 3.323546 | 1.82 | 0.069 | -.4643753 | 12.56368 |
| lnhouse | -.5208677 | .1841968 | -2.83 | 0.005 | -.8818869 | -.1598485 |
| lnpop | .9008388 | .1420921 | 6.34 | 0.000 | .6223434 | 1.179334 |
| _cons | -13.20082 | 2.922405 | -4.52 | 0.000 | -18.92863 | -7.473013 |